\documentclass[referee]{raa}            % referee version: for submission

\usepackage{graphicx,times}             %for PS/EPS graphics inclusion, new
\usepackage{natbib}
\usepackage{amssymb,amsmath}
\bibpunct{(}{)}{;}{a}{}{,}

\usepackage[a4paper=true,dvipdfm=true,pagebackref=true]{hyperref}
\hypersetup{colorlinks = true, linkcolor = green, anchorcolor = red, citecolor = blue, filecolor = red, pagecolor = red, urlcolor = red}

\begin{document}

   \title{Position and attitude determination by integrated GPS/SINS/TS for feed support system of FAST
\footnotetext{$*$ Corresponding author.}
}
%   \subtitle{I. Place Your Subtitle Here}

   \volnopage{Vol.0 (20xx) No.0, 000--000}      %%preserved for Editor. DOn't remove!
   \setcounter{page}{1}          %%starting page, preserved for Editor. DOn't remove!

   \author{Ming-Hui Li
      \inst{1,2}
   \and Peng Jiang
      \inst{1*}
   \and Dong-Jun Yu
      \inst{1}
  \and Jing-Hai Sun
      \inst{1}
   }

   \institute{National Astronomical Observatories, Chinese Academy of Sciences,
             Beijing 100012, China; {\it liminghui@nao.cas.cn;pjiang@nao.cas.cn }\\
        \and
             University of Chinese Academy of Sciences, Beijing 100049, China
\\
%        \and
%             Full institute address for the third author\\
\and
%             Full institute address for the fourth author\\
\vs\no
   {\small Received~~2019 Dec 17; accepted~~2020~~Apr 8}}

\abstract{In this paper, a new measurement system based on integration method is presented, which can provide all-weather dependability and higher precision for the measurement of FAST$'$s feed support system. The measurement system consists of three types of measuring equipments, and a processing software with the core data fusion algorithm. The Strapdown Inertial Navigation System(SINS) can autonomously measure the position, speed and attitude of the carrier. Its own shortcoming is the measurement data diverges rapidly over time. SINS must combine the Global Positioning System(GPS) and the Total Station(TS) to obtain high-precision measurement data. Kalman filtering algorithm is adopted for the integration measurement system, which is an optimal algorithm to estimate the measurement errors. To evaluate the performance, series of tests are carried out. For the feed cabin, the maximum RMS of the position is 14.56mm, the maximum RMS of the attitude is 0.095$\dg$, these value are less than 15mm and 0.1$\dg$ as the precision for measuring the feed cabin. For the Stewart manipulator, the maximum RMS of the position is 2.99mm, the maximum RMS of the attitude is 0.093$\dg$, these value are less than 3mm and 0.1$\dg$ as the precision for measuring the Stewart manipulator. As a result, the new measurement meets the requirement of measurement precision for FAST$'$s feed support system.
}

   \authorrunning{Ming-Hui Li, Peng Jiang, Dong-Jun Yu, Jing-Hai Sun}            %author_head in even pages
   \titlerunning{Position and attitude determination by integrated GPS/SINS/TS for feed support system of FAST}  % title_head in odd pages

   \maketitle

%
%________________________________________________ sections below
%
\section{Introduction}           %% first-level sections will be auto-capitalized
\label{sect:intro}

The Five-hundred-meter Aperture Spherical radio Telescope (FAST) is the world$'$s largest single dish and the most sensitive radio telescope located in Guizhou Province, China. The construction was completed in September 2016. After 3 years$'$ commissioning work, the telescope is now ready for open running.

The orientation of the telescope receiver is controlled by the feed support system. The telescope$'$s feeds are mounted in a 30-ton feed cabin at a height of about 140 meters and with a movement range of about 206 meters. The feed cabin is suspended and driven by six parallel cables. To realize high-accurate positioning of the feeds, a fine-adjusting mechanism including an AB rotator and a Stewart manipulator is inserted between feeds and the cabin body to compensate the vibration and deviation caused by large-span cables. The real-time position and attitude of the feed cabin are measured and compared with the theoretical planning value. The difference value is provided to the control mechanism of six parallel cables as control information. In order to accurately locate the feeds, the real-time position and attitude of Stewart manipulator need to be measured to provide a basis for adjusting the Stewart manipulator.

According to the sensitivity requirements of the telescope, the pointing accuracy of the telescope receiver is required to be 16$''$, the control accuracy of the feed cabin is 30mm. The measurement accuracy of the feed-cabin$'$s position  is 15mm, the measurement accuracy of its attitude is about 0.1$\dg$, the measurement accuracy of the Stewart manipulator$'$s position is 3mm, and the measurement accuracy of its attitude is about 0.1$\dg$(\citealt{Peng Jiang+etal+2019}).

In the early stage of FAST commissioning, the measurement scheme of the feed support system adopts total station. The angle accuracy of total station can reach 0.5$''$, the position accuracy of short-range can reach less than 0.5mm. However, total station has the disadvantages of uncertain time delay and is sensitive to the weather condition. In the weather of rain, fog and light intensity, the total station fails to find the target and the measurement accuracy decreases significantly. This decrease affects the control accuracy of the feed support system and results in the pointing deviation of the telescope feeds. When the deviation is too large, the telescope fail to observe.

In order to extend the effective observation time of FAST, the measurement scheme of the feed support system should meet  all-weather requirement. The total station cannot meet this demand, so the measurement scheme is upgraded to the integration measurement. The new measurement scheme includes strapdown inertial navigation system(SINS), global positioning system(GPS) and total station(TS). SINS provides position, speed and attitude data without  external information, but SINS can only guarantee short-term accuracy. However, SINS works independently for a long time, the measurement data diverges rapidly and the measurement accuracy decreases significantly(\citealt{Qin Y Y+2014}).The current RealTime Kinematic(RTK) GPS can achieve centimetre-level position accuracy with a local reference station. It is well known that multiple GPS antennas can be used to determine the attitude(\citealt{Crassidis J L+etal+1997}). GPS has shortcomings such as satellite signal loss, whole-cycle ambiguity, and signal multipath effect. The integrated SINS/GPS can make up these shortcomings and provide all-weather and high-accurate measurement data about position and attitude. The measurement of the feed cabin uses integrated GPS/SINS. The position measurement accuracy of the Stewart manipulator is 3mm, only TS can meet this high demand, so the Stewart manipulator uses integrated TS/SINS. The integration algorithm utilizes Kalman filter, which has became a standard data fusion approach in GPS/SINS integration system. Kalman filter is also used for the integrated TS/SINS on the Stewart manipulator.

\begin{figure}
   \centering
   \includegraphics[width=10cm]{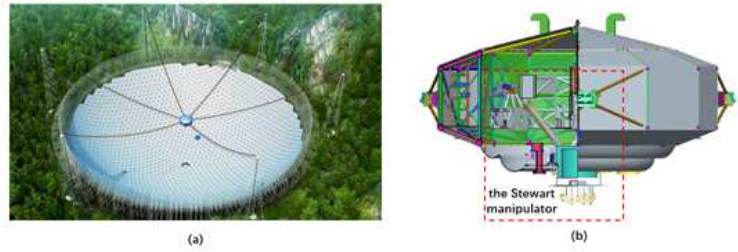}
   \caption{{\small (a) Illustration of the FAST telescope, (b) The feed cabin and Stewart manipulator.} }
   \label{Fig1}
   \end{figure}

\section{Description of Integration Measurement Scheme}
\label{sect:Obs}

The tracking speed of the feed cabin is 22mm/s, and the maximum speed of the changing source is 400mm/s.The maximum inclination of the feed cabin is 15$\dg$, and the maximum inclination of the Stewart manipulator is 30$\dg$. As shown in Figure 2(a), on the outer frame of the feed cabin, six choke antennas are installed at equal intervals to receive satellite signals, among which three are beidou receivers and the other three are leica receivers. They are mutually backup to improve the system reliability. Two local reference stations are installed on the mountain near the telescope, and their position coordinates are known accurately. RTK was used to obtain the center position of the feed cabin. Two high-precision strapdown inertial navigation system devices are installed in the feed cabin, which one for the feed cabin measurement and the other for the Stewart manipulator measurement. SINS device consists of 3-axis gyroscope, 3-axis accelerometer. Six high-precision total station instruments are placed on the measurement stations through FAST$'$s reflecting surface, to measure position of the Stewart manipulator by the distance intersection method in Figure 2(b).

\begin{figure}
   \centering
   \includegraphics[width=10cm, angle=0]{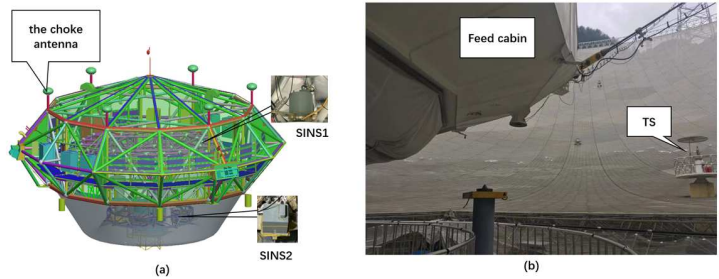}
   \caption{{\small Illustration of the hardware in the integration measurement system.} }
   \end{figure}

\subsection{Hardware Specifications}
In order to ensure high-precision position and attitude measurement, the integration measurement system uses high-precision measurement devices such as SINS, GPS receiver and TS . The error distribution of each device is shown in Table 1 to 3:

\begin{table}
\begin{center}
\caption[]{ SINS performance }\label{Tab:SINS gyro}
 \begin{tabular}{clcl}
  \hline\noalign{\smallskip}
 Index name      & gyro value  & acc value                    \\
  \hline\noalign{\smallskip}
 Scale factor repeatability & 20ppm  & 100ppm \\ % new variable
 Scale factor nonlinearity    &20ppm  & 100ppm \\
 Measurement range     & $\pm$400$\dg$/s  & $\pm$50g  \\
 Stability of zero offset & 0.01$\dg$/h(1$\sigma$)  & 100$\mu$g \\
 Repeatability of zero offset & 0.01$\dg$/h(1$\sigma$)  & 100$\mu$g \\
 Random walk  & 0.001$\dg$/h$^{\rm 1/2}$   & N/A \\
  \noalign{\smallskip}\hline
\end{tabular}
\end{center}
\end{table}

\begin{table}
\begin{center}
\caption[]{GPS receiver performance}\label{Tab:GPS receiver performance}
 \begin{tabular}{clcl}
  \hline\noalign{\smallskip}
 Index name      & Value                     \\
  \hline\noalign{\smallskip}
 Plane static accuracy & $\pm$6mm    \\ % new variable
 Altitude static accuracy & $\pm$8.5mm   \\
 Combined postprocessing accuracy  &  $\pm$5mm+1ppm    \\
 Data update frequency & 20Hz \\
  \noalign{\smallskip}\hline
\end{tabular}
\end{center}
\end{table}

\begin{table}
\begin{center}
\caption[]{TS performance}\label{Tab:TS performance}
 \begin{tabular}{clcl}
  \hline\noalign{\smallskip}
 Index name      & Value                     \\
  \hline\noalign{\smallskip}
 Angle error & 0.5$''$\\
 Position error  &0.5mm+1ppm    \\ % new variable
 Time delay & 10ms   \\
 Measure rang & 1.5m--3500m\\
 Measure time & 2.4s(one time)\\
  \noalign{\smallskip}\hline
\end{tabular}
\end{center}
\end{table}

\subsection{Measurement Process}

The process of the measurement system is shown in Figure 3. The integration measurement system starts, TS initializes to find the measurement targets and receives the real-time data from the weather station. GPS receivers search for the available satellite signals and establish a communication link with the local reference stations. SINS conducts initial alignment to determine the initial attitude of the measured object, i.e. yaw, pitch and roll. Then the SINS updates the attitude, speed and position, and finally Kalman filter is used for information integration to obtain the high-accurate measurement results. Because the feed support system moves at a low speed, the yaw error calculated by SINS is large, and a multiple GPS antennas scheme is adopted to determine the yaw of the feed cabin. Both the A-B axis and the Stewart manipulator are rigid bodies, so the yaw of the Stewart manipulator can be derived by extrapolating attitude of the GPS/SINS integration system.When TS can not use, the position and attitude of the Stewart manipulator are calculated with the feed cabin's measurement information and the moving value of A-B axis and six bars in the feed cabin. The A-B axis and six bars are controlled by the optical encoder multiturn absolute, this encoder's attitude accuracy is less than 0.4$''$, its position accuracy is less than 1mm. This precision can meet the requirement, such as attitude accuracy is 0.1$\dg$ and position accuracy is 3mm, for measuring the Stewart manipulator.
\begin{figure}
   \centering
   \includegraphics[width=10cm, angle=0]{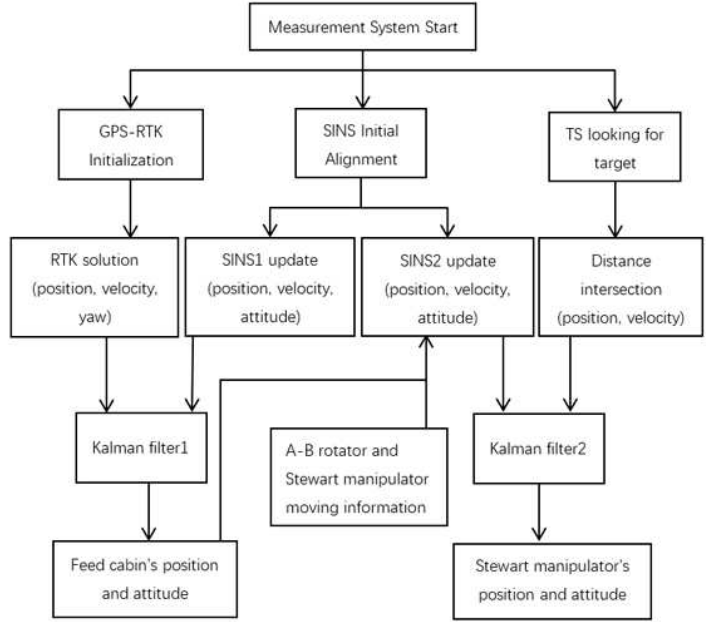}
   \caption{Process of the integration measurement system. }
   \label{Fig3}
   \end{figure}

\section{Calculation Method of Integrated Measurement}
\subsection{SINS Initial Alignment}
The initial alignment of SINS is a process of determining the reference navigation coordinate frame. SINS is just startup, the orientation of the feed-cabin body coordinate frame($b$-frame), relativing to each axis of the reference navigation coordinate frame($n$-frame), is completely unknown or not accurate enough. Therefore, it cannot enter the navigation state immediately. The coarse alignment method can determine the angle between $b$-frame and $n^{\prime}$-frame with two vectors such as $g^{b}$ and $\omega_{i e}^{b}$. When the feed support system is stationary or moving at a constant speed, the gyros measure the similar earth rotation rate, the accelerometers measure the local gravitational acceleration(\citealt{Gu D+etal+2008}).

The attitude matrix $C_{b}^{n^{\prime}}$, which relates the body frame($b$-frame) to the computational navigation frame($n^{\prime}$-frame), could be calculated by the following equation:
\begin{equation}
  C_{b}^{n^{\prime}}=\left[\begin{array}{c}{\left(g^{n}\right)^{T}} \\ {\left(\omega_{i e}^{n}\right)^{T}} \\ {\left(g^{n} \times \omega_{i e}^{n}\right)^{T}}\end{array}\right]^{-1}\left[\begin{array}{c}{\left(g^{b}\right)^{T}} \\ {\left(\omega_{i e}^{b}\right)^{T}} \\ {\left(g^{b} \times \omega_{i e}^{b}\right)^{T}}\end{array}\right]
\label{eq:DVO}
\end{equation}

\begin{equation}
\begin{array}{l}{\tilde{\omega}_{i b}^{b} \approx \omega_{i e}^{b}} \\
{\tilde{f}^{b} \approx-g^{b}}\end{array}
\end{equation}
where $g^{n}$ is the gravity vector in $n$ frame, $\tilde{f}^{b}$ is specific force vector from accelerometers output. $\omega_{i e}^{n}$ is the earth rotation rate resolved in $n$ frame, $\tilde{\omega}_{i b}^{b}$ is angular rate vector from gyro output.

After coarse alignment, this is a fine alignment. It calculates the misalignment angle($\phi$) between  $n^{\prime}$-frame and $n$-frame. The attitude matrix $C_{b}^{n^{\prime}}$ is modified by $\phi$ to get a more accurate attitude matrix $C_{b}^{n}$ :
\begin{equation}
C_{b}^{n}=C_{n^{\prime}}^{n} C_{b}^{n^{\prime}}=[I+(\phi \times)] C_{b}^{n^{\prime}}
\end{equation}

\subsection{SINS Numerical Update}
The updating algorithm of SINS consists of position-updating, velocity-updating and attitude-updating. The attitude -updating is the core, and its precision plays a decisive role in the accuracy of the SINS.
The attitude-updating can be calculated as follows:
\begin{equation}
\dot{\boldsymbol{C}}_{b}^{n}=\boldsymbol{C}_{b}^{n}\left(\boldsymbol{\omega}_{n b}^{b} \times\right)
\end{equation}
The velocity-updating can be described as follows:
\begin{equation}
\dot{V}^{n}=C_{b}^{n} f^{b}-\left(2 \omega_{i e}^{n}+\omega_{e n}^{n}\right) \times V^{n}+g^{n}
\end{equation}
The position-updating can be calculated as follows:
\begin{equation}
\dot{r}^{n}=V^{n}-{\omega}_{e n}^{n} \times r^{n}
\end{equation}
where superscript $n$ refers to the $n$-frame, superscript $b$ refers to the $b$-frame. $C_{b}^{n}$ is attitude matrix, $\dot{\boldsymbol{C}}_{b}^{n}$ is attitude rate matrix. $\boldsymbol{\omega}_{n b}^{b}$ is the angular rate vector of $b$-frame with respect to $n$-frame projected in $b$-frame , $\omega_{e n}^{n}$ is the angular rate vector of $n$-frame with respect to earth frame ($e$-frame) projected in $n$-frame. $V^{n}$ is velocity vector, $\dot{V}^{n}$ is velocity rate vector, $\mathbf{r}^{n}$ is position vector, $\dot{\mathbf{r}}^{n}$ is position rate vector.

By simultaneous equations(4) to (6), the attitude, velocity and position of the carrier can be calculated. However, the SINS has inherent defects such as divergence of navigation accuracy over time and poor long-term stability. To solve this problem, the measurement system aids the GPS and TS devices.
\subsection{Kalman Filter}
The essence of integration measurement is state estimation. The Kalman filter is a set of mathematical equations
that provides an efficient computational means to estimate the state of a process, and minimizes the root mean squared error of that state(\citealt{Welch G+2001}). The state vector of Kalman filter is 15 dimensions in the integration measurement system, is described as follows:
\begin{equation}
\begin{aligned} \mathbf{X}=&\left[\phi_{E}, \phi_{N}, \phi_{U}, \delta V_{E}, \delta V_{N}, \delta V_{U}\right.\\ &\left.\delta L, \delta \lambda, \delta h, \varepsilon_{x}, \varepsilon_{y}, \varepsilon_{z}, \nabla_{x}, \nabla_{y}, \nabla_{z}\right]^{T} \end{aligned}
\end{equation}
Where $E$, $N$ and $U$ are respectively east, north and up. The last six terms are the gyro zero drift error and the accelerometer sensor error, which are repeatability errors of inertial devices that have great impact on the accuracy of the system. The first nine terms are,respectively, attitude, velocity, and position error vectors, which can be derived from (4) to (6).\\
The derivation of attitude error vectors is as follows:
\begin{equation}
\omega_{n b}^{b}=\omega_{i b}^{b}-\omega_{i n}^{b}
\end{equation}
Substitute (8) and (3) into (4) and use the attitude quaternion method, and obtain the attitude error vector equation (9).
\begin{equation}
\dot{\phi}=\phi \times \omega_{i n}^{n}+\delta \omega_{i n}^{n}-C_{b}^{n}\left(\left[\delta K_{G}\right]+[\delta G]\right) \omega_{i b}^{b}-\varepsilon^{n}\\
\end{equation}
The derivation of velocity error vectors is as follows:\\
Equation (5) is the ideal velocity equation, and the actual velocity equation is as follows:
\begin{equation}
\dot{\tilde{V}}^{n}=\tilde{C}_{b}^{n} \tilde{f}^{b}-\left(2 \tilde{\omega}_{i e}^{n}+\tilde{\omega}_{e n}^{n}\right) \times \tilde{V}^{n}+\tilde{g}^{n}
\end{equation}
Subtract equation (10) from equation (5)
\begin{equation}
\begin{aligned} \delta \dot{V}^{n} &=\left[(I-\phi \times) C_{b}^{n}\left(f^{b}+\delta f^{b}\right)-C_{b}^{n} f^{b}\right] \\
&-\left\{\left[2\left(\omega_{i e}^{n}+\delta \omega_{i e}^{n}\right)+\left(\omega_{e n}^{n}+\delta \omega_{e n}^{n}\right)\right] \times\left(V^{n}+\delta V^{n}\right)-\left(2 \omega_{i e}^{n}+\omega_{e n}^{n}\right) \times V^{n}\right\}+\delta g^{n} \end{aligned}
\end{equation}
Expand (11) and omit the second order small quantities of the error, the velocity error vectors equation (12):
\begin{equation}
\delta \dot{V}^{n}=-\phi^{n} \times f^{n}+C_{b}^{n}\left(\left[\delta K_{A}\right]+[\delta A]\right) f^{b}+\delta V^{n} \times\left(2 \omega_{i e}^{n}+\omega_{e n}^{n}\right)+V^{n} \times\left(2 \delta \omega_{i e}^{n}+\delta \omega_{e n}^{n}\right)+\nabla^{n}\\
\end{equation}
The derivation of position error vectors is as follows:
\begin{equation}
\omega_{e n}^{n}=\left[\begin{array}{c}-\frac{V_{N}}{R_{M}+h} \\ \frac{V_{E}}{R_{N}+h} \\ \frac{V_{E}}{R_{N}+h} \tan L\end{array}\right]
\end{equation}
Substitute the angular rate of b-frame $\omega_{e n}^{n}$ into equation (6), and solve the differential equation to get the position vectors, they are as follows:
\begin{equation}
\dot{L}=\frac{1}{R_{M}+h} V_{N}, \quad \dot{\lambda}=\frac{\sec L}{R_{N}+h} V_{E}, \quad \dot{h}=V_{U}
\end{equation}
Take the derivatives of equations (14), and get equations (15) - (17):
\begin{equation}
\delta \dot{L}=\delta V_{N} /\left(R_{M}+h\right)-\delta h V_{N} /\left(R_{M}+h\right)^{2}
\end{equation}
\begin{equation}
\delta \dot{\lambda}=\delta V_{E} /\left(R_{N}+h\right) \sec L+\delta L \tan L \sec L V_{E} /\left(R_{N}+h\right)-\delta h V_{E} \sec L /\left(R_{N}+h\right)^{2}
\end{equation}
\begin{equation}
\delta \dot{h}=\delta V_{U}
\end{equation}
where $L$ is latitude, $\lambda$ is longitude, $h$ is height. $K_{G}$ is the gyro scale coefficient error, $G$ is the gyro installation error. $K_{A}$ is the accelerometer scale coefficient error, $A$ is the accelerometer installation error. $R_{M}$ and $R_{N}$ are constants.

In integration measurement system, the measurement error is often used as a state vector. Because the Kalman filter is a linear filter, the prediction equation of the measurement error is linear. Generally, the measurement error is relatively small, its higher order items can be ignored to simplify the prediction equation of the measurement error(\citealt{Yan G M+2007}). The Kalman filter equation includes state equation and measurement equation, which are as follows:
\begin{equation}
X_{k}=F X_{k-1}+W_{k-1}
\end{equation}
\begin{equation}
Z_{k}=H X_{k}+V_{k}
\end{equation}
where subscript ${k-1}$ is the last time, $k$ is the current time. $F$ is the state transition matrix, $H$ is the measurement matrix, $W$ is the system noise, $V$ is the measurement noise, $Z$ is a measurement vector, in the integration of GPS and SINS, the form of measurement vector $Z_{G-S}$ is as equation (20), and in the integration of TS and SINS, the form of measurement vector $Z_{T-S}$ is as equation (21).
\begin{equation}
Z_{G-S}=\left[V_{G E}-V_{S E}, V_{G N}-V_{S N}, V_{G U}-V_{S U}, L_{G}-L_{S}, \lambda_{G}-\lambda_{S}, h_{G}-h_{S}\right]^{T}
\end{equation}
\begin{equation}
Z_{T-S}=\left[V_{T E}-V_{S E}, V_{T N}-V_{S N}, V_{T U}-V_{S U}, L_{T}-L_{S}, \lambda_{T}-\lambda_{S}, h_{T}-h_{S}\right]^{T}
\end{equation}

In practice, due to the strong observability of position and velocity vector, the initial value of state vector and corresponding root mean square error matrix is allowed to be relatively large, and they will converge rapidly with the update of filtering. Secondly, the smaller the variance matrix of system noise is, the lower the utilization rate of the measurement vector is. However, the smaller the variance matrix of measurement noise is, the higher the utilization rate of measurement vector is, and vice versa(\citealt{Yan G M+2019}). Kalman filter automatically adjusts the utilization rate of state information and measurement information according to the size of state noise and measurement noise, so as to make the most reasonable estimation of the current state. In the process of filtering, the optimal estimation value of state vector is continuously used to modify the SINS calculation value, to make it close to the real position and attitude value. The feedback of Kalman filter is helpful to keep the measurement error equation linear.
\section{Data analysis}
\label{sect:analysis}
The experiment has been conducted on FAST, in Guizhou province, China, to evaluate the integration measurement system. In the experiment, the total stations and the astronomical observation trajectory were used as the references to evaluate the accuracy of the integration measurement solution. The experiment was carried out when the weather conditions is available without rain, fog and strong sunshine, which ensures stable operation of TS. The telescope changes the source in first 180 seconds of the experiment, to lock onto the observed objects, followed by about 600 seconds of tracking the objects. Because the integration measurement$'$s accuracy of the tracking source is the main focus, the analyzing of tracking  source is described in detail. The analyzing process of changing source is similar to the tracking source, so only Table 6 is used to show the integration measurement$'$s accuracy in the changing source state.The experiment$'$s source is 0029+349, the observation time is from 15:32:10 to 15:45:02 on February 25, 2020. The right ascension of this source is 00:29:14.24, its declination is 34$\dg$56$'$32.2$''$. In tracking state, the zenith angle is [11.5$\dg$,12.9$\dg$], the azimuth is [319.4$\dg$,326.6$\dg$].
\subsection{Verify Accuracy of GPS/SINS Results}
The GPS/SINS measurement scheme is used for the feed cabin. The measurement data of total station can be used as a reference. The total station$'$s attitude measurement accuracy is 0.5$''$, the position measurement accuracy is 5mm, which are all superior to the measurement requirements of the feed cabin. At the outermost edge of the feed cabin, there are six optical targets. The total station can track these six targets to work out the real-time position and attitude of the feed cabin, which can be used as a reference value to verify the accuracy of GPS/SINS results.\\
In the accuracy experiment of GPS/SINS, since the difference between TS data and GPS/SINS integrated data is a very small value compared with the experimental data range, TS data and GPS/SINS integrated data cannot be differentiated in a figure. Therefor, only the integrated GPS/SINS result are shown in Figure 4 and 5, the error result between GPS/SINS data and TS data are shown in Figure 6 and 7.
%% tracking for feed cabin
\begin{figure}
   \centering
   \includegraphics[width=10cm, angle=0]{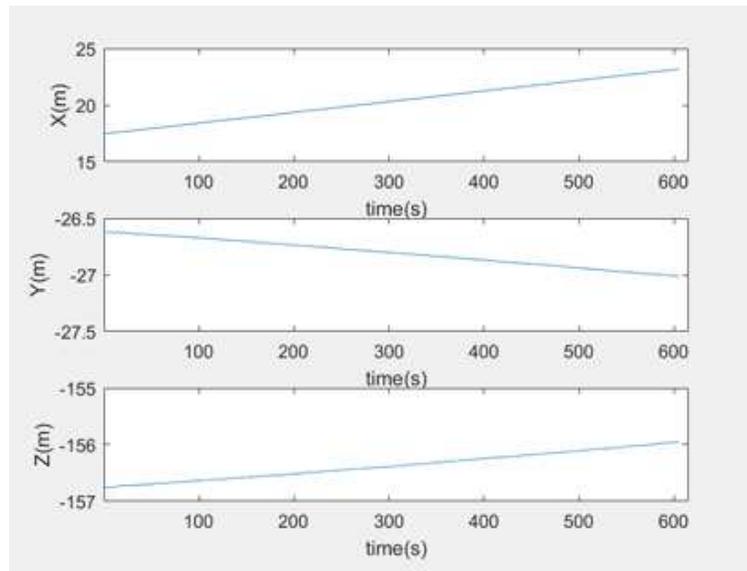}
   \caption{Integrated position of the feed cabin during tracking source.}
   \label{Fig4}
   \end{figure}

\begin{figure}
   \centering
   \includegraphics[width=10cm, angle=0]{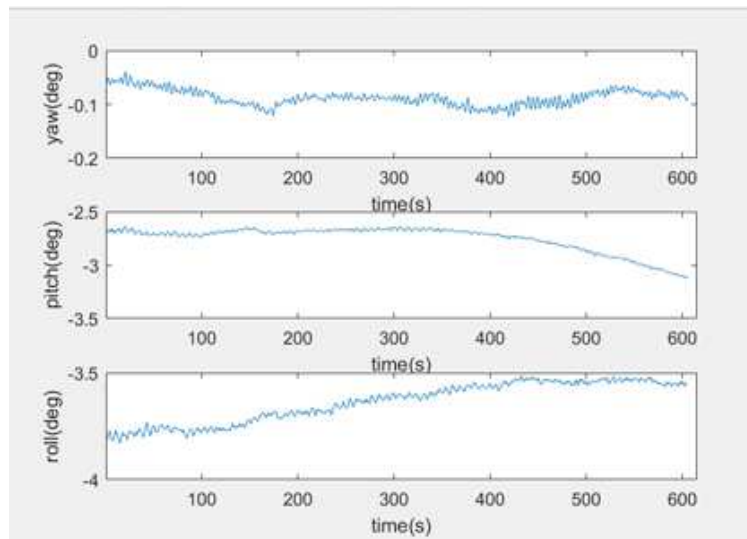}
   \caption{Integrated attitude of the feed cabin during tracking source.}
   \label{Fig5}
   \end{figure}

\begin{figure}
   \centering
   \includegraphics[width=10cm, angle=0]{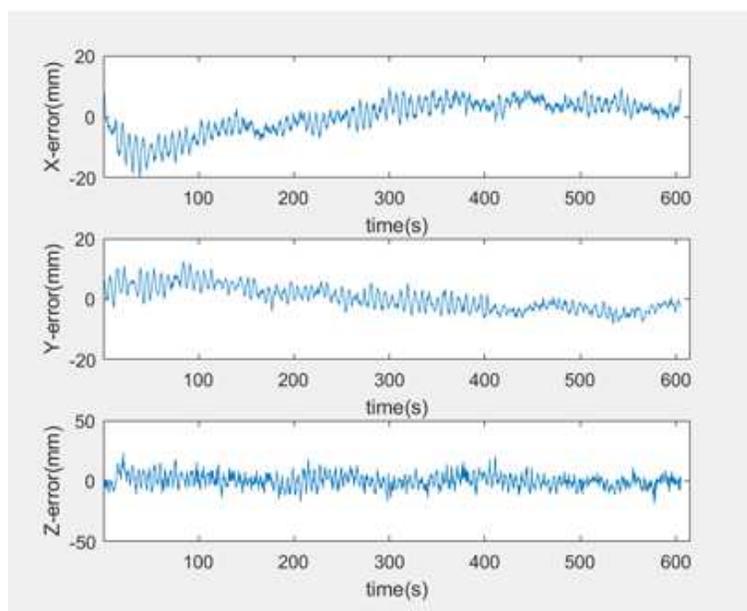}
   \caption{Position error between GPS/SINS and TS during tracking source.}
   \label{Fig6}
   \end{figure}

\begin{figure}
   \centering
   \includegraphics[width=10cm, angle=0]{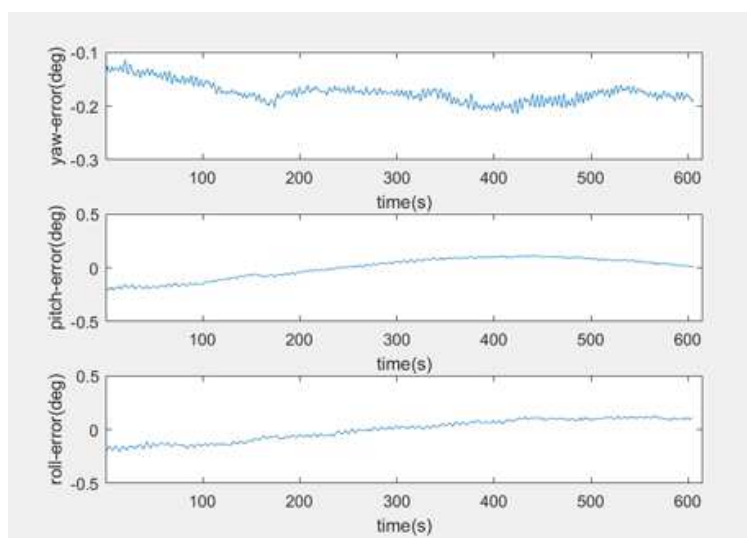}
   \caption{Attitude error between GPS/SINS and TS during tracking source.}
   \label{Fig7}
   \end{figure}

\begin{table}
\bc
\begin{minipage}[]{100mm}
\caption[]{The error between GPS/SINS and TS in tracking.\label{tab1}}\end{minipage}
\setlength{\tabcolsep}{1pt}
\small
 \begin{tabular}{ccccccccccccc}
  \hline\noalign{\smallskip}
Feed cabin    &    X &     Y &    Z  &   yaw &   pitch  &  roll\\
  \hline\noalign{\smallskip}
Tracking Ave &-0.11mm	&0.27mm	&2.84mm &-0.175$\dg$	&-0.002$\dg$	&-0.001$\dg$ \\
Tracking Rms &5.58mm	&4.05mm	&5.62mm &0.018$\dg$	    &0.095$\dg$	&0.095$\dg$ \\
  \noalign{\smallskip}\hline
\end{tabular}
\ec
\end{table}
From Table 4, the maximum root mean squared errors of the angle is 0.095$\dg$, the maximum root mean squared errors of the position is 5.62mm. These value are less than 15mm and 0.1$\dg$ as the precision for measuring the feed cabin, so the position and attitude accuracy of integrated GPS/SINS can meet the measurement requirements of the feed cabin.
\subsection{Verify Accuracy of TS/SINS Results}
The TS/SINS measurement scheme is used for the Stewart manipulator. According to the design requirements, the position measurement accuracy of the Stewart manipulator is 3mm. However, in a scale of 206 meters and a height of 140 meters, there is no measurement equipment with a higher precision than 3mm. No specific measuring equipment can be found as a reference. From the perspective of practical application, the main task of the Stewart manipulator is making telescope receivers to accurately accept the radio source signal. Therefore, the astronomical observation trajectory is used as a reference for the measurement accuracy of the Stewart manipulator.

According to the right ascension, declination and observation time of the source 0029+349, the astronomical observation trajectory (azimuth and zenith angle) in the horizontal coordinate system can be calculated, it is shown in Figure 8. With this value, the planned position trajectory (X-Y-Z) of Stewart manipulator in the FAST coordinate system can be derived, then the planned attitude trajectory (yaw-pitch-roll) of Stewart manipulator can be calculated according to the deformation strategy of the telescope. By comparing the error between the planned trajectory of the Stewart manipulator and the TS/SINS integrated measurement, to verify accuracy of TS/SINS Results. The planned trajectory of the Stewart manipulator and the integrated TS/SINS result of tracking source are shown in Figure 9 and 10. In a zooming of Figure 9, there is the red line is planned position trajectory of the Stewart manipulator and the blue line is TS/SINS integrated position. In Figure 10, there is the red line is planned attitude trajectory of the Stewart manipulator and the blue line is TS/SINS integrated attitude. The error result between the planned trajectory and TS/SINS data are shown in Figure 11 and 12.
\begin{figure}
   \centering
   \includegraphics[width=10cm, angle=0]{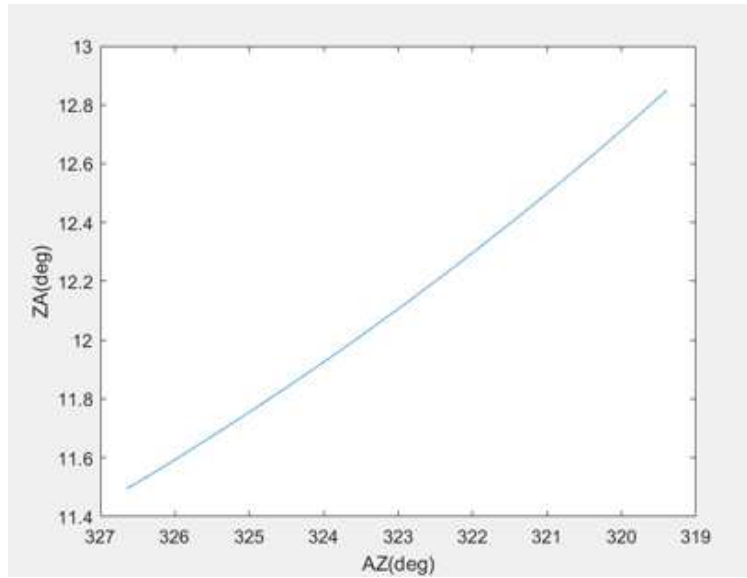}
   \caption{Astronomical observation trajectory in the horizontal coordinate.}
   \label{Fig8}
   \end{figure}
%% tracking for the Stewart manipulator
\begin{figure}
   \centering
   \includegraphics[width=10cm, angle=0]{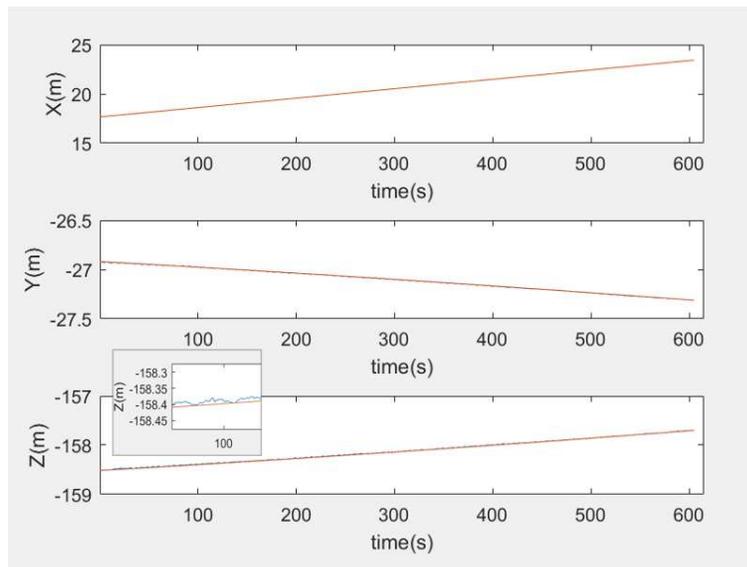}
   \caption{Position of planned trajectory and TS/SINS during tracking source.}
   \label{Fig9}
   \end{figure}

\begin{figure}
   \centering
   \includegraphics[width=10cm, angle=0]{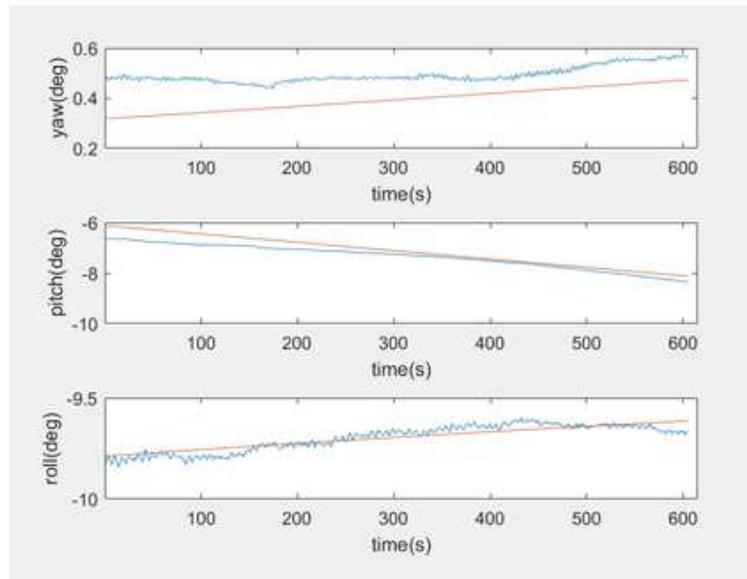}
   \caption{Attitude of planned trajectory and TS/SINS during tracking source.}
   \label{Fig10}
   \end{figure}

\begin{figure}
   \centering
   \includegraphics[width=10cm, angle=0]{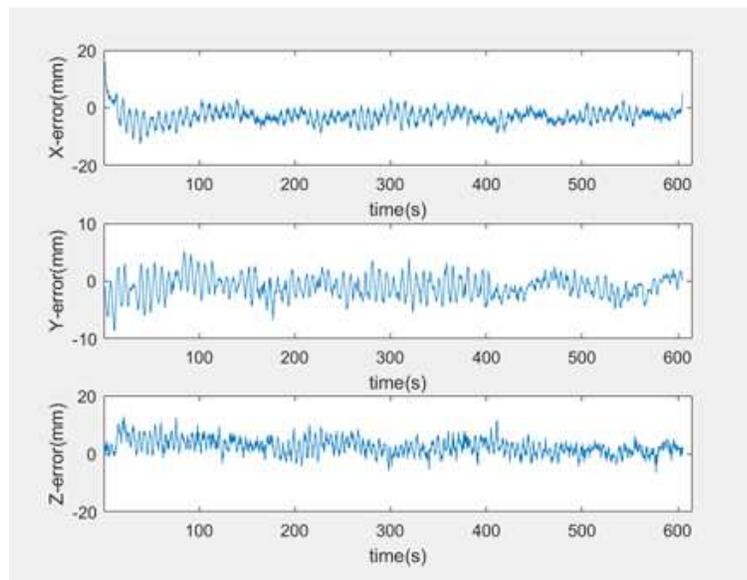}
   \caption{Position error between planned trajectory and TS/SINS in tracking.}
   \label{Fig11}
   \end{figure}

\begin{figure}
   \centering
   \includegraphics[width=10cm, angle=0]{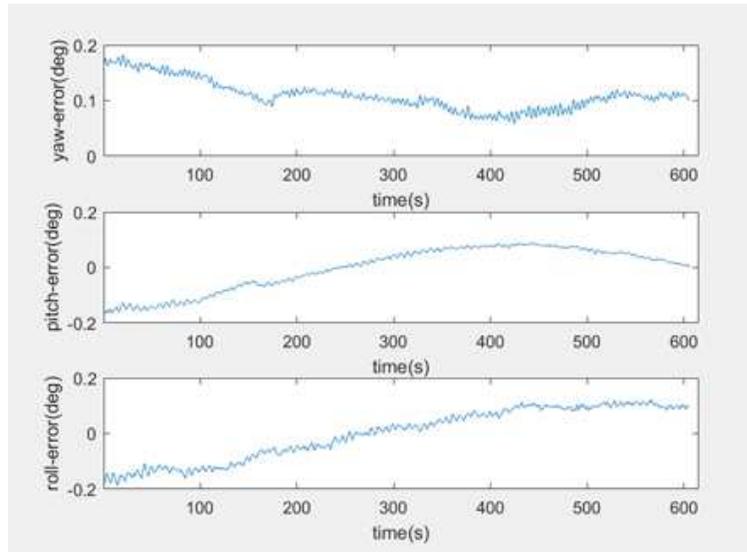}
   \caption{Attitude error between planned trajectory and TS/SINS in tracking.}
   \label{Fig12}
   \end{figure}

\begin{table}
\bc
\begin{minipage}[]{100mm}
\caption[]{The error between planned trajectory and TS/SINS in tracking.\label{tab1}}\end{minipage}
\setlength{\tabcolsep}{1pt}
\small
 \begin{tabular}{ccccccccccccc}
  \hline\noalign{\smallskip}
Stewart manipulator    &    X &     Y &    Z  &   yaw &   pitch  &  roll\\
  \hline\noalign{\smallskip}
Tracking Ave &-2.82mm	&-1.00mm	&2.48mm &0.110$\dg$	&-0.004$\dg$	&-0.001$\dg$ \\
Tracking Rms &2.62mm	&1.91mm	&2.76mm &0.027$\dg$	    &0.078$\dg$	&0.093$\dg$ \\
  \noalign{\smallskip}\hline
\end{tabular}
\ec
\end{table}
From Table 5, the maximum root mean squared errors of the angle is 0.093$\dg$, the maximum root mean squared errors of the position is 2.76mm. These value are less than 3mm and 0.1$\dg$ as the precision for measuring the Stewart manipulator, so the position and attitude accuracy of integrated TS/SINS can meet the measurement requirements of the Stewart manipulator.
\begin{table}
\bc
\begin{minipage}[]{100mm}
\caption[]{The error between reference and measuring in changing source.\label{tab1}}\end{minipage}
\setlength{\tabcolsep}{1pt}
\small
 \begin{tabular}{ccccccccccccc}
  \hline\noalign{\smallskip}
various error  &    X &     Y &    Z  &   yaw &   pitch  &  roll\\
  \hline\noalign{\smallskip}
Feed cabin Ave &-0.96mm	 &6.72mm	&2.90mm  &-0.120$\dg$	&-0.141$\dg$	&-0.137$\dg$\\
Feed cabin Rms &14.56mm	&6.43mm	&6.12mm  &0.088$\dg$	&0.094$\dg$	 &0.061$\dg$	    \\
Stewart manipulator Ave  &-0.62mm	 &-0.13mm	&4.26mm  &0.078$\dg$	&0.006$\dg$	&-0.172$\dg$\\
Stewart manipulator Rms &2.99mm	&2.48mm	&2.99mm  &0.049$\dg$	&0.093$\dg$	 &0.075$\dg$	    \\
  \noalign{\smallskip}\hline
\end{tabular}
\ec
\end{table}

In Table 6, the error between reference and measuring in changing source are shown. For the feed cabin in the changing source state, the maximum root mean squared errors of the angle is 0.094$\dg$, the maximum root mean squared errors of the position is 14.56mm. For the Stewart manipulator in the changing source state, the maximum root mean squared errors of the angle is 0.093$\dg$, the maximum root mean squared errors of the position is 2.99mm. So the position and attitude accuracy of integrated measurement can meet the requirements in changing source state.
\section{Conclusions}
\label{sect:conclusion}
The integration measurement system can provide all-weather dependability and higher precision for the measurement of FAST$'$s feed support system. The Kalman filter algorithm is used to integrate SINS data with GPS data to generate the attitude and position solutions of the feed cabin. Meanwhile, the algorithm fuses SINS data with TS data to generate the attitude and position solutions of the Stewart manipulator. To ensure a non-divergence and available solution in the long term, the Kalman filter algorithm constantly updates the estimation of measurement errors. The TS and the astronomical trajectory are used as the references to evaluate the accuracy of the integration measurement solution. The experimental results show that the integration measurement system is stable and can correctly determine attitude and position of the feed support system. The RMS meets precision requirements.

\begin{acknowledgements}
This work was funded by the National Natural Science Foundation of China (Grant No.11673039), the Open Project Program of the Key Laboratory of FAST, NAOC, Chinese Academy of Sciences and the Key Laboratory of Radio Astronomy, Chinese Academy of Sciences.
\end{acknowledgements}

\label{lastpage}

\end{document}